\documentstyle[12pt]{article}

\begin{document}
\newcommand \s[1] {\mbox{\hspace{#1cm}}}
\newcommand \FR[2]
 {\s{.06}{\strut\displaystyle#1\over\strut\displaystyle#2}\s{.06}}
\newcommand \fr[2] {\s{.06}{\textstyle{#1\over#2}}\s{.06}}
\newcommand \eqnum[1]{\eqno (#1)}
\renewcommand \d {{\rm d}}
\hyphenation{counter-term counter-terms}

\thispagestyle{empty}
\setcounter{page}0

\begin{flushright}
JINR-E2-94-388
\end{flushright}
\vglue 1cm

\begin{center}

{}~\vfill

{\large \bf THE BACKGROUND-FIELD METHOD AND \vspace{4mm} \protect \\
 NONINVARIANT RENORMALIZATION\footnote{Supported in part by ISF grant \#
 RFL000}}

\vfill

{\large L. V. Avdeev\footnote{E-mail: $avdeevL@thsun1.jinr.dubna.su$},
 D. I. Kazakov\footnote{E-mail: $kazakovD@thsun1.jinr.dubna.su$}}
and
{\large M. Yu. Kalmykov\footnote{E-mail:
 $kalmykov@thsun1.jinr.dubna.su$}}

\vspace{2cm}

{\em Bogoliubov Laboratory of Theoretical Physics,\\
 Joint Institute for Nuclear Research,\\
 $141\,980$ Dubna $($Moscow Region$)$, Russian Federation}

\end{center}

\vfill

\begin{abstract}

We investigate the consistency of the back\-ground-field formalism when
applying various regularizations and renormalization schemes. By an
example of a two-di\-men\-sion\-al $\sigma$ model it is demonstrated
that the back\-ground-field method gives incorrect results when the
regularization (and/or renormalization) is noninvariant. In particular,
it is found that the cut-off regularization and the differential
renormalization belong to this class and are incompatible with the
back\-ground-field method in theories with nonlinear symmetries.

\end{abstract}

\vfill

\pagebreak

\section{Introduction} \indent

To obtain meaningful results in quantum field theory, one has to remove
ultraviolet and infrared divergencies. This goal can be achieved by a
renormalization procedure, that is, a proper subtraction of
singularities. In a general case, ultraviolet renormalization involves
three steps:

\begin{enumerate}

\item Regularization of Feynman amplitudes by introducing some parameter
which converts divergencies into singularities as this parameter tends
to a particular limit value (say, zero or infinity). There should exist
a smooth limit of taking the regularization off: any amplitudes that
were finite without it should not be distorted.

\item Renormalization of the parameters of the theory (coupling
constants, masses,etc) in order to absorb the singularities
into a redefinition of these parameters, which is achieved by
introducing some local counterterms.

\item The choice of a renormalization scheme which fixes the finite
arbitrariness left after the regularization is taken off.

\end{enumerate}

In gauge theories, one has to be very careful because the
renormalization procedure may violate the gauge invariance on the
quantum level, thus destroying the renormalizability of the theory.
Therefore, when dealing with gauge theories, one is bound to apply an
invariant renormalization. By this we mean a renormalization that
preserves all the relevant symmetries of the model on the quantum level,
that is, preserves all the Ward identities \cite{Sl} for the
renormalized Green functions.

On the one hand, this can be achieved by applying an invariant
regularization first (respecting the symmetries in the regularized
theory) and then using, for instance, the minimal subtraction scheme
\cite{min,Sp} to fix the finite arbitrariness. On the other hand, when
the regularization is noninvariant or no explicit regularization is
introduced at all, there is no automatic preservation of the symmetries.
Then one has to take care of that directly. For example, one can use
some noninvariant regularization and consecutively choose certain finite
counterterms to restore the invariance (the symmetry of the renormalized
Green functions) order by order in perturbation theory~\cite{B}.

When treating theories with nonlinear realization of a symmetry, like
two-di\-men\-sion\-al $\sigma$ models or quantum gravity, one faces
extraordinary complexity of perturbative calculations. To simplify them,
one usually applies the so-called back\-ground-field method \cite{back}
which allows one to handle all the calculations in a strictly covariant
way. This method was successfully applied to multiloop calculations in
various gauge and scalar models, being combined with the minimal
subtraction scheme based on some invariant regularization.

The most popular and handy regularization used in these calculations was
the dimensional regularization \cite{dim}. It has been proved to be an
invariant regularization, preserving all the symmetries of the classical
action that do not depend explicitly on the space-time dimension
\cite{Sp,action}. Moreover, any formal manipulations with the
dimensionally regularized integrals are allowed. However, an obvious
drawback of this regularization is the violation of the axial invariance
and of supersymmetry. That is why there are numerous attempts to find
some other regularization equally convenient and efficient. Among such
schemes the recently proposed differential renormalization \cite{diff}
is discussed.

In the present paper we investigate the compatibility of the
back\-ground-field formalism with various regularizations and
renormalization prescriptions. Our conclusion is that the
back\-ground-field method necessarily requires one to use an invariant
renormalization procedure. As the invariance does not hold, the method
gives incorrect results. We demonstrate this by an example of the
two-di\-men\-sion\-al $O(n)$ $\sigma$-model, comparing the dimensional
regularization, the cut-off regularization within the minimal
subtraction scheme, and the differential renormalization method.

\section{Invariant Renormalization in the Back\-ground-Field Formalism}
\indent

To preserve all the symmetries on the quantum level, one has to apply an
invariant renormalization procedure. The simplest way to construct such
a procedure is to use an invariant regularization and the minimal
subtraction scheme. An invariant regularization should permit any formal
manipulations with the functional integral that are needed to ensure the
Ward identities. Let us list the properties of an invariant
regularization~\cite{invariant}:

\begin{enumerate}
\item translational invariance $\int\d^D x~ f(x+y)=\int\d^D x~ f(x)$;

\item unambiguity of the order of integrations $\int$$\d^D x $$\int$$
\d^D y$ $f(x,y)$=$\int$$\d^D y$$\int$$\d^D x$ $f(x,y)$;

\item linearity
$\int\d^D x \sum_j a_j$ $f_j(x)$=$\sum_j a_j \int\d^D x$ $f_j(x)$;

\item Lorentz covariance;

\item integration by parts, neglecting the surface terms;

\item possibility of canceling the numerator with the denominator;

\item commutativity of the space-time or momentum integration and
differentiation with respect to an external parameter.

\end{enumerate}

The only known regularization obeying all these requirements is the
dimensional regularization. Combined with the minimal subtraction
scheme, it makes an invariant renormalization for which the action
principle is valid \cite{action}. Using the dimensional renormalization
in conjunction with the back\-ground-field method leads to covariant
results of multiloop calculations in any theory unless its symmetry
properties depend on the particular number of dimensions.

On the other hand, if one applies some noninvariant regularization, the
initial symmetry may be violated, and one has to explore the possibility
of using such a regularization in the framework of the
back\-ground-field method.

In multiloop calculations by this method an invariant regularization
automatically provides some implicit correlations between different
diagrams, which may be essential as the formal back\-ground-field
expansion of the action is performed. Any violation of these fine
correlations by a noninvariant regularization or by an improper choice
of finite counterterms may result in wrong answers, although covariant
in form.

\section{Two-dimensional nonlinear $O(n)~~ \sigma $ model}
\indent

Let us consider the two-di\-men\-sion\-al $\sigma$ model of the $O(n)$
principal chiral field (${\bf n}$ field) and calculate the two-loop
$\beta$ function, using various approaches. The model is described by
the lagrangian

\begin{equation} \label{n-field} {\cal L} = \FR 1 {2h}
 \left( \partial_\mu {\bf n} \right) ^2, \s2 {\bf n}^2 = 1.
\end{equation}

\noindent It can be treated as a special case of the generic bosonic
$\sigma$ model

\begin{equation} \label{sigma}
 {\cal L} = \fr 1 2 (\partial_\mu \phi^j)~ g_{jk}(\phi)~
 \partial_\mu \phi^k , \s2 (j,k=1,2,...,n-1) ,
\end{equation}
where the metric is of the form
\begin{equation} \label{metric}
 g_{jk}(\phi)= \delta_{jk} +\FR {h~ \phi_j\phi_k} {1-h~\phi^2} .
\end{equation}

The back\-ground-field expansion of the action can be done in a strictly
covariant fashion~\cite{sigma}.

To separate the ultraviolet and infrared divergencies, we add an
auxiliary mass term to the initial lagrangian~(\ref{sigma})

\begin{equation} \label{mass}
 {\cal L}_m= \fr 1 2 m^2~ \phi^j~ g_{jk}(\phi)~ \phi^k .
\end{equation}

\noindent This additional term serves only for eliminating infrared
divergencies, naively present in any two-di\-men\-sion\-al theory with
massless scalars. After the calculation of the ultraviolet logarithms,
one should set $m^2$=0.

The $\sigma$ model (\ref{sigma}) with the particular choice of the
metric (\ref{metric}) becomes renormalizable. All the covariant
structures that may appear as counterterms are reducible to the metric,
so that the only thing that happens is a renormalization of the kinetic
term. By rescaling the fields the renormalization can be absorbed into
the charge. The invariant charge $\overline{h}$=$Z^{-1}h$ is defined
through the field renormalization constant $Z$. To calculate $Z$ within
the back\-ground-field method, one has to consider the
one-particle-irreducible diagrams with two external lines of the
background field, and quantum fields inside the loops. Up to two loops
the relevant diagrams \cite{Bel} are shown in fig.~1. Their
contributions to $Z$ are obtained by normalizing to the tree term
$\Bigl[
-$$\fr 1 2 (\partial_\mu \phi^j)$ $g_{jk}$ $\partial_\mu \phi^k
\Bigr]$.

\begin{figure}[htbp] 

 $$ \FR 1 2 R_{jk}~ (\partial_\mu \phi^j)~ (\partial_\mu \phi^k) \s{.4}
  \begin{picture}(10,7)(-5,-2.5) 
   \put(0,0){\circle 7}
   \put(0,-3.5){\line(-1,-1){3.6}}
   \put(0,-3.5){\line(1,-1){3.6}}
   \put(0,-3.5){\circle*1}
  \end{picture} \eqnum{\rm a} $$

 $$ - \FR 1 {12}
  \left( 2~ {R^a}_j~ R_{ak} + 3~ {R^{abc}}_j~ R_{abck}
  \right) (\partial_\mu \phi^j)~ (\partial_\mu \phi^k) \s{.4}
  \begin{picture}(10,7)(-5,-1) 
   \put(0,2.8){\circle{5.6}}
   \put(0,-2.8){\circle{5.6}}
   \put(-4,0){\line(1,0)8}
   \put(0,0){\circle*1}
  \end{picture} \eqnum{\rm b} $$

 $$ + \FR 1 6 {R_j}^{ab}{}_k~ R_{ab}~
  (\partial_\mu \phi^j)~ (\partial_\mu \phi^k) \s{.4}
  \begin{picture}(10,8)(-5,-1) 
   \put(0,0){\circle{5.6}}
   \put(0,5.6){\circle{5.6}}
   \put(-2.8,0){\vector(0,-1){.7}} 
   \put(2.8,0){\vector(0,-1){.7}}
   \put(0,0){\circle*{.6}}
   \put(0,2.8){\circle*1} 
   \put(0,-2.8){\circle*1}
   \put(0,-2.8){\line(-1,-1){3.6}} 
   \put(0,-2.8){\line(1,-1){3.6}}
  \end{picture} \eqnum{\rm c} $$

 $$ - \FR 4 9 {R_j}^{(ab)c}~ R_{kabc}~
  (\partial_\mu \phi^j)~ (\partial_\nu \phi^k) \s{.4}
  \begin{picture}(16,9)(-8,-1) 
   \put(0,0){\circle{9.8}}
   \put(-7,0){\line(1,0){14}}
   \put(-4.9,0){\circle*1}
   \put(4.9,0){\circle*1}
   \put(-2.5,0){\vector(1,0){.7}} 
   \put(2.5,0){\vector(-1,0){.7}}
  \end{picture} \eqnum\d $$

 $$ - \FR 8 9 {R_j}^{(ab)c}~ R_{k(cb)a}~
  (\partial_\mu \phi^j)~ (\partial_\nu \phi^k) \s{.4}
  \begin{picture}(16,8)(-8,-1) 
   \put(0,0){\circle{9.8}}
   \put(-7,0){\line(1,0){14}}
   \put(-4.9,0){\circle*1}
   \put(4.9,0){\circle*1}
   \put(-3.5,3.5){\vector(1,1){.5}} 
   \put(3.5,-3.5){\vector(-1,-1){.5}}
  \end{picture} \eqnum{\rm e} $$

 \caption{The one- and two-loop corrections to the effective action of
  the two-di\-men\-sion\-al bosonic $\sigma$ model without torsion.
  Lines of the diagrams refer to propagators $1/(p^2$+$m^2)$, and arrows
  to $p_\mu$ in numerators.}

\end{figure}

The Riemann and Ricci tensors are the functionals of the background
field. In our model with the metric given by eq.~(\ref{metric}) they are
evaluated to

$$ R_{abjk}= h \left( g_{aj}~g_{bk} -g_{ak}~g_{bj} \right) , \s2
 R_{jk}=(n-2)~h~g_{jk}~. $$

Besides, we should take into account the renormalization of the mass
$Z_{m^2}$. In the first loop it is determined by the diagram of fig.~2
(normalized to $\Bigl[-\fr 1 2 m^2~ \phi^j~ g_{jk}~ \phi^k \Bigr]$).
Although this operator gives no direct contribution to the wave-func\-tion
renormalization, in all the diagrams that contribute to $Z$ the mass
ought to be shifted by such corrections. In the two-loop approximation,
only the fig.~2 correction to fig.~1 is essential.

\begin{figure}[htbp] 
 $$ \FR 1 6 m^2~ R_{jk}~ \phi^j \phi^k \s{.4}
  \begin{picture}(10,7)(-5,-2.5) 
   \put(0,0){\circle 7}
   \put(0,-3.5){\line(-1,-1){3.6}}
   \put(0,-3.5){\line(1,-1){3.6}}
   \put(0,-3.5){\circle*1}
  \end{picture} $$
 \caption{The one-loop mass correction to the effective action.}
\end{figure}

The two-loop renormalizations can be carried out either directly --- via
a certain ${\cal R}$ operation, diagram by diagram, --- or by means of
re-expanding the one-loop counterterms in the background field and the
quantum field (provided we have an intermediate regularization, and the
counterterms can be written down explicitly). The additional diagrams
that emerge in this way are shown in fig.~3.

\begin{figure}[htbp] 

 $$ \FR 1 2 R_{aj}~ R^a{}_k~ (\partial_\mu \phi^j)~
  (\partial_\mu \phi^k)~ (-{\cal K R}'~{\rm fig.~1a}) \s{.4}
  \begin{picture}(10,6)(-5,-2.5) 
   \put(0,0){\circle 7}
   \put(0,-3.5){\line(-1,-1){3.6}}
   \put(0,-3.5){\line(1,-1){3.6}}
   \put(0,-3.5){\circle 2}
   \put(0,-2.5){\line(0,-1)2}
  \end{picture} \eqnum{\rm a} $$

 $$ -\FR 1 2 R_{jabk}~ R^{ab}~ (\partial_\mu \phi^j)~
  (\partial_\mu \phi^k)~ (-{\cal K R}'~{\rm fig.~1a}) \s{.4}
  \begin{picture}(10,6)(-5,-2.5) 
   \put(0,0){\circle 7}
   \put(0,-3.5){\line(-1,-1){3.6}}
   \put(0,-3.5){\line(1,-1){3.6}}
   \put(-3.5,0){\vector(0,-1){.7}} 
   \put(3.5,0){\vector(0,-1){.7}}
   \put(0,0){\circle*{.6}}
   \put(0,3.5){\circle 2}
   \put(0,2.5){\line(0,1)2}
   \put(0,-3.5){\circle*1}
  \end{picture} \eqnum{\rm b} $$

 $$ \FR 1 2 m^2~ R_{jabk}~ R^{ab}~ (\partial_\mu \phi^j)~
  (\partial_\mu \phi^k)~ (-{\cal K R}'~{\rm fig.~2}) \s{.4}
  \begin{picture}(10,6)(-5,-2.5) 
   \put(0,0){\circle 7}
   \put(0,-3.5){\line(-1,-1){3.6}}
   \put(0,-3.5){\line(1,-1){3.6}}
   \put(0,-3.5){\circle*1}
   \put(0,3.5){\circle 2}
   \put(0,2.5){\line(0,1)2}
  \end{picture} \eqnum{\rm c} $$

 \caption{The diagrams emerging from the back\-ground-field expansion of
  the one-loop counterterms.}

\end{figure}

One can find the $\beta$ function by requiring independence of the
invariant charge on the normalization point. This leads to the following
expression for the $\beta$ function through the finite wave-function
renormalization constant:

\begin{equation} \label{beta} \beta (h)= h
 \left( \mu^2 \FR {\partial Z} {\partial \mu^2}
 \right)
 \left[ \Bigl( 1-h \FR \partial {\partial h} \Bigr) Z
 \right] ^{-1} .
\end{equation}

Using the dimensional regularization and the minimal subtraction scheme
to calculate the diagrams presented above, one obtains the following
well-known expression for the two-loop $\beta$ function of the ${\bf
n}$-field model (\ref{n-field}) \cite{n-field,Bel}:

\begin{equation} \label{beta_dim}
 \beta_{\rm dim}= -(n-2) \FR {h^2} {4\pi}
 \Bigl( 1 +2 \FR h {4\pi} \Bigr) .
\end{equation}

As it has already been mentioned, the dimensional regularization and the
minimal subtraction scheme provide us with an invariant renormalization
procedure within the back\-ground-field method. Hence, the obtained
expression for the $\beta$ function is correct, and we can use it as a
reference expression to compare with other approaches. Owing to the
presence of just one coupling constant in the model, the $\beta$
function should be renormalization-scheme independent up to two loops
and should coincide with eq.~(\ref{beta_dim}).

To check the validity of the back\-ground-field method in conjunction
with other regularizations and renormalization prescriptions, let us
consider the calculation of the $\beta$ function within two schemes:
the cut-off regularization and the differential renormalization.

\subsection{The Cut-Off Regularization} \indent

We start with the regularization that uses a cut-off in the momentum
space. All the integrals over the radial variable in the Euclidean space
are cut at an upper limit~$\Lambda$. Strictly speaking, this is not a
very promising regularization, since it explicitly breaks the Lorentz,
as well as gauge, invariance. However, we use it here to realize what
may happen when a noninvariant regularization is applied.

We are going to use the minimal subtraction scheme which respects the
invariance properties of the applied intermediate regularization, keeps
them intact, as they are.

When the regularization parameter has the dimension of a mass, the
minimal subtraction procedure can be defined \cite{log} so as just to
convert the logarithms of the (infinite) cut-off~$\Lambda$ into the
logarithms of a finite renormalization point $\mu$ which appears in the
theory after renormalizations:

\begin{equation} \label{K}
 {\cal K}~ \ln^n (\Lambda^2)= \ln^n (\Lambda^2) -\ln^n (\mu^2),
 \s2 {\cal K}~ \Lambda^n = \Lambda^n ,
\end{equation}
so that
\begin{equation} \label{R}
 {\cal R}~ \ln^n (\Lambda^2) = (1-{\cal K})~ \ln^n (\Lambda^2) =
 \ln^n (\mu^2), \s2 {\cal R}~ \Lambda^n = 0.
\end{equation}

\noindent In case of overlapping divergencies, which generate powers of
the logarithms, one ought to perform the standard renormalization
procedure prior to the subtractions. However, if only the final
renormalized answers are of interest, one can simply drop all the
contributions of the minimally subtracted counterterms (\ref{K}), since
they will be annihilated by (1$-$${\cal K}$), eq.~(\ref{R}),
irrespective of any powers of the logarithms from the residual graphs
with contracted subgraphs. The same will happen to all the diagrams of
fig.~3, generated by re-expanding the counterterms.

Thus, it is sufficient to calculate the regularized diagrams of figs.~1
and 2, up to $\Lambda$-power corrections and ultra\-violet-finite
two-loop contributions, and then to replace $\Lambda^2 $ by $\mu^2 $ and
$m^2$ in the one-loop diagram of fig.~1a by $m^2 Z_{m^2}$, including
the correction from fig.~2. The contributions of individual diagrams are

\begin{eqnarray*}
Z({\rm fig.~1a}) &=& -(n-2)~ h/(4\pi)~ \ln(\mu^2/m^2), \\
Z({\rm fig.~1b}) &=& \fr 1 3 (n-2)~(n+1)~ h^2/(4\pi)^2~
 \ln^2 (\mu^2/m^2), \\
Z({\rm fig.~1c}) &=& -\fr 1 3 (n-2)^2~ h^2/(4\pi)^2
 \left[ \ln^2 (\mu^2/m^2) -\ln(\mu^2/m^2) \right] , \\
Z({\rm fig.~1d}) &=& -\fr 2 3 (n-2)~ h^2/(4\pi)^2~
 \ln^2 (\mu^2/m^2), \\
Z({\rm fig.~1e}) &=& -\fr 1 3 (n-2)~ h^2/(4\pi)^2~
 \ln^2 (\mu^2/m^2), \\
Z_{m^2}({\rm fig.~2}) &=& -\fr 1 3 (n-2)~ h/(4\pi)~ \ln(\mu^2/m^2).
\end{eqnarray*}
The charge-renormalization constant proves then to be
\begin{equation} \label{Z_cut}
 Z_{\rm cut}= 1 -(n-2) \FR h {4\pi} \ln \FR {\mu^2} {m^2} +0\cdot h^2,
\end{equation}
so that eq.~(\ref{beta}) gives the $\beta$ function
\begin{equation} \label{beta_cut}
 \beta_{\rm cut}(h)= -(n-2) \FR {h^2} {4\pi}
 \left( 1+ 0\cdot h \right) .
\end{equation}

The difference between this result and that obtained in dimensional
renormalization (\ref{beta_dim}) is a direct manifestation of the
noninvariance of the cut-off regularization, which violates the
translational invariance. However, one needs to explain the reason for
the failure to reproduce the correct $\beta$ function in the present
case. Although the cut-off regularization is noninvariant, still it has
been successfully used to perform multiloop calculations in scalar field
theories and in the quantum electrodynamics both within the
back\-ground-field method and by the conventional diagram technique.

The point is that those theories were renormalizable in the ordinary
sense, that is, they had a finite number of types of divergent diagrams.
In contrast, the ${\bf n}$-field model is renormalizable only in the
generalized sense. The total number of divergent structures here (with
various external lines) is infinite, but they are related to each other
by general covariance of the renormalized theory (in case of an
invariant renormalization). So the number of independent structures
remains finite. Expanding the lagrangian, we get an infinite number of
terms; however, the renormalization constants are not arbitrary but
mutually related. Although the back\-ground-field method formally
preserves the covariance of the model, the use of a noninvariant
renormalization would break the intrinsic connection between various
diagrams (and between their renormalization constants), thus leading to
wrong results.

Therefore, we conclude that in generalized renormalizable (as well as
nonrenormalizable) theories it is not allowed to use the cut-off
regularization with the minimal subtractions in the framework of the
back\-ground-field method.

We now want in the same way to check the invariance properties of the
differential renormalization method.

\subsection{Differential Renormalization} \indent

The idea of the differential renormalization traces back to the
foundations of the renormalization procedure \cite{Bog} as a
redefinition of the product of distributions at a singular point. The
method suggests to work in the co-ordinate space, where the free Green
functions are well defined, although their product at coinciding points
suffers from ultraviolet divergencies. The divergencies manifest
themselves as singular functions which have no well-defined Fourier
transform. The recipe of the differential renormalization \cite{diff}
consists in rewriting a singular product in the form of a differential
operator applied to a nonsingular expression:

\begin{equation} \label{rewrite}
 f(x_j,...,x_k)= D
 \bigl( \Box^{\sigma_j}_{x_j},...,\Box^{\sigma_k}_{x_k} \bigr)~
 g(x_j,...,x_k) ,
\end{equation}

\noindent Eq.~(\ref{rewrite}) should be understood in the sense of
distributions, that is, in the sense of integration with a test
function. Then one ignores any surface terms on rearranging the
derivatives via integration by parts. The nonsingular function
$g(x_j,...,x_k)$ is obtained by solving a differential equation, and
hence, involves an obvious arbitrariness. The latter can be identified
with the choice of a renormalization point and a renormalization scheme.
In this respect the differential renormalization does not differ from
any other renormalization prescription.

In the absence of a primary regularization this prescription might
preserve all the needed invariances and, what is important for
applications, seems to renormalize ultraviolet singularities in the
integer dimension. On the other hand, the absence of any intermediate
regularization prevents one from using the standard scheme: invariant
regularization + minimal subtractions. Therefore, to verify the
invariance properties of the differential renormalization, one has to
deal with renormalized amplitudes directly.

Two-loop calculations of the renormalization constant in the
two-di\-men\-sion\-al $\sigma$ model in the framework of the
differential renormalization have been performed in
ref.~\cite{troubles}. The authors have used the concept of the infrared
$\widetilde{\cal R}$ operation to handle the infrared divergencies. In
the present case the infra\-red-re\-norm\-al\-ized free propagator in
the co-ordinate representation has the form

\begin{equation} \label{R_tilde}
 \widetilde{\cal R}~ \Delta_0(x)= -\FR 1 {4\pi} \ln(x^2 N^2),
\end{equation}
where $N^2$ is an infrared renormalization scale.

An important role in the calculations plays the tadpole diagram
(fig.~1a). In four dimensions, diagrams of this type diverge
quadratically and can be consistently renormalized to zero, as it was
originally done in the method of the differential renormalization
\cite{diff}. However, in two dimensions the leading one-loop
contribution to the $\beta$ function comes from this very diagram.
Hence, the tadpole should be different from zero in any renormalization.
This means that we have to define the two-di\-men\-sion\-al tadpole
diagram in a self-con\-sist\-ent way in addition to the recipe of the
differential renormalization. Such an extension has been discussed in
detail in ref.~\cite{troubles}, where the following expression for the
massless tadpole has been suggested:

\begin{equation} \label{R*}
{\cal R}^{*} \Delta_0(0)= \FR 1 {4\pi} \ln \FR {M^2} {N^2}.
\end{equation}

\noindent The parameter $M^2$ is an ultraviolet scale, and ${\cal R}^*$
denotes the complete infrared and ultraviolet renormalization.

According to this modification of the differential renormalization
rules, the expression for the $\beta$ function has been found to be

\begin{equation} \label{beta_diff0}
 \beta_{\rm diff}= -(n-2) \FR {h^2} {4\pi} +0\cdot h^3 .
\end{equation}

\noindent Thus, the definition of the tadpole via eq.~(\ref{R*}) in the
massless case gives the correct expression for the {\em one}-loop
$\beta$ function, but fails in two loops.

Therefore, we would like to circumvent possible ambiguities of combining
the infrared $\widetilde{\cal R}$ operation with the differential
renormalization. We are going to apply the method to the massive model
in which no infrared difficulties ever appear.

Introducing the mass term according to eq.~(\ref{mass}), we obtain the
free propagator of the form

\begin{equation} \label{D_m} \Delta_m (x)= \FR 1 {2\pi} K_0(m |x|),
\end{equation}
where $K_0$ is the MacDonald function, obeying the equation
\begin{equation} \label{K_eq}
 (\partial^2-m^2)~ K_0(m|x|)= -2\pi~\delta_{(2)}(x).
\end{equation}
Bearing in mind the known expansion of the MacDonald function
$$
K_0(x)=-\ln \Bigl( \FR x 2 \Bigr) \sum_{k=0}^{\infty} \FR 1 {(k!)^2}
 \Bigl( \FR x 2 \Bigr) ^{2k} + \sum_{k=0}^{\infty}
 \FR {\psi(k+1)} {(k!)^2} \Bigl( \FR x 2 \Bigr) ^{2k} ,
$$

\noindent we come to the following natural generalization of
eq.~(\ref{R*}) to the massive case:

\begin{equation} \label{tad1}
 {\cal R} \Bigl[ \delta(x)~ \Delta_m (x) \Bigr] = \delta(x)
 \FR 1 {4\pi} \ln \FR {M^2} {m^2} .
\end{equation}

In due course of the calculation we shall also need to define the
product of two tadpoles. In the spirit of the consistent ${\cal R}$
operation, the squared tadpole (fig.~1b) should be defined as the square
of the renormalized value (\ref{tad1}) for fig.~1a, that is,

\begin{equation} \label{tad2}
 {\cal R} \Bigl[ \delta(x)~ \Delta_m^2 (x) \Bigr] = \delta(x)
 \left( \FR 1 {4\pi} \ln \FR {M^2} {m^2} \right) ^2 .
\end{equation}

Now we are in a position to complete the calculation of the $\beta$
function. We present it in more detail for the diagrams of fig.~1c and
fig.~1d. The simple tadpole subgraph of fig.~1c has already been defined
via eq.~(\ref{tad1}). Consider another subgraph, with the numerator,

\begin{equation} \label{tad1p}  -\int\d^2 y
 \Bigl[ \FR \partial {\partial y_\nu} \Delta_m (x-y) \Bigr] ^2 .
\end{equation}

\noindent Integrating by parts, ignoring the surface term, and then
using eq.~(\ref{K_eq}), we get

$$ \int\d^2 y
 \Bigl[ m^2 \Delta_m^2 (x-y) -\delta(x-y)~ \Delta_m (x-y)
 \Bigr] . $$

The integral of the first term is finite and known (the normalization
integral for $K_0$). On the other hand, the second term is just reduced
to the basic tadpole (\ref{tad1}). Thus, the result for
eq.~(\ref{tad1p}) is $1/(4\pi)~ [1 -\ln(M^2/m^2)]$.

Generally speaking, the systematic differential renormalization to all
orders \cite{all} allows for introducing different ultraviolet
renormalization scales in different diagrams (all the scales varying
proportionally to each other under re\-norm\-al\-iz\-a\-tion-group
transformations). The ratio of these parameters can then be specially
chosen \cite{diff} to satisfy the Ward identities. Let us denote the
scale that appears in the tadpole with the numerator by $M_1$.

Proceed now to fig.~1d. Its contribution to the effective action is
$$ \FR 1 3 \int \d^2 x \int \d^2 y~ R^{abc}{}_j(x)~ R_{abck}(y)
 \left[ \partial_\mu \phi^j(x) \right]
 \left[ \partial_\nu \phi^k(y) \right] \Delta_m^2 (x-y)~
 \partial_\mu \partial_\nu \Delta_m (x-y). $$

\noindent Picking out the trace and traceless parts according to
ref.~\cite{diff}, we get

\pagebreak[3]
$$ \FR 1 3 \int \d^2 x \int \d^2 y~ R^{abc}{}_j(x)~ R_{abck}(y)
 \left[ \partial_\mu \phi^j(x) \right]
 \left[ \partial_\nu \phi^k(y) \right] \Delta_m^2 (x-y) \times $$
$$
 \Bigl[
  \left( \partial_\mu \partial_\nu -\fr 1 2 \delta_{\mu\nu} \partial^2
  \right) \Delta_m (x-y)
  +\fr 1 2 m^2 \delta_{\mu\nu} \Delta_m (x-y)
  +\fr 1 2 \delta_{\mu\nu} \left( \partial^2-m^2 \right) \Delta_m (x-y)
 \Bigr] . $$

\noindent The first term, which is traceless, is finite and does not
generate any ultraviolet scale; one can easily establish this fact in
the momentum representation. The second term vanishes as $m^2$$\to$0.
Thus, we are left with the last term. Via eq.~(\ref{K_eq}) it is reduced
to eq.~(\ref{tad2}), that is, gives only the square of the logarithm.
However, again the renormalization scale $M_2$ in the new diagram may
differ from $M$.

Below we present the contributions of all the diagrams to the
renormalization constants:

\begin{eqnarray*}
Z({\rm fig.~1a}) &=& -(n-2)~ h/(4\pi)~ \ln(M^2/m^2), \\
Z({\rm fig.~1b}) &=& \fr 1 3 (n-2)~(n+1)~ h^2/(4\pi)^2~
 \ln^2 (M^2/m^2), \\
Z({\rm fig.~1c}) &=& -\fr 1 3 (n-2)^2~ h^2/(4\pi)^2~ \ln(M^2/m^2)
 \left[ \ln (M_1^2/m^2) -1 \right] , \\
Z({\rm fig.~1d}) &=& -\fr 2 3 (n-2)~ h^2/(4\pi)^2~ \ln^2 (M_2^2/m^2), \\
Z({\rm fig.~1e}) &=& -\fr 1 3 (n-2)~ h^2/(4\pi)^2~ \ln^2 (M_3^2/m^2), \\
Z_{m^2}({\rm fig.~2}) &=& -\fr 1 3 (n-2)~ h/(4\pi)~ \ln(M^2/m^2).
\end{eqnarray*}
This gives the $\beta$ function
\begin{equation} \label{beta_diff}
 \beta_{\rm diff}= -(n-2) \FR {h^2} {4\pi}
 \left\{ 1 +\FR h {4\pi}
  \left[ \FR 1 3 (n-2)~ \ln \FR {M_1^2} {M^2}
   +\FR 4 3 \ln \FR {M_2^2} {M^2} +\FR 2 3 \ln \FR {M_3^2} {M^2}
  \right]
 \right\} .
\end{equation}

\noindent We see that the result explicitly depends on the ratio of the
renormalization scale parameters in different diagrams. Such a
dependence on the details of the renormalization prescription is beyond
the usual scheme arbitrariness. It would never occur to two loops in the
conventional perturbation theory for ordinary renormalizable one-charge
models. There the arbitrariness would be completely absorbed into a
finite number of counterterms which are of the operator types present in
the tree lagrangian. Hence, we should try to fix the parameters of the
differential renormalization by imposing some additional requirements.
In the quantum electrodynamics the gauge Ward identities could be used
to this end \cite{diff}. For the $\sigma$ model in the
back\-ground-field formalism the situation is not so clear.

The parameter $M_1$ that appears in the one-loop tadpole subgraph of
fig.~1c with the numerator can be fixed as follows. In the momentum
representation we can easily see that the sum of this diagram and the
simple tadpole (fig.~1a) is just an ultra\-violet-finite integral which
equals $1/(4\pi)$. The value will be correctly reproduced by the
differential renormalization if we choose the same scale for both
tadpole graphs: $M_1$=$M$. Thus, for these diagrams the renormalization
seems to be automatically invariant.

Let us point out that this simple check is by no means trivial. For
example, the straightforward Feynman regularization of the quantum-field
propagator in the momentum space $1/(p^2+m^2) \to 1/(p^2+m^2)
-1/(p^2+M^2)$ would not stand the test. As a result, the coefficient of
the lower-order logarithm generated by fig.~1c would be incorrect, and a
contribution proportional to $(n-2)^2$ would be left in the two-loop
$\beta$ function [as for $M_1$$\ne$$M$ in eq.~(\ref{beta_diff})]. The
Feynman cut-off is therefore a noninvariant regularization and cannot be
freely combined with the back\-ground-field method.

Expecting that the differential renormalization is automatically
invariant, we would set $M_2$=$M_3$=$M$ as well. However, then
eq.~(\ref{beta_diff}) would again give us the wrong result
(\ref{beta_diff0}) obtained under the assumption of the automatic
invariance in the massless theory via the infrared $\widetilde{\cal R}$
operation. Hence, the ratio of the renormalization parameters ought to
be somehow tuned in order to restore the invariance.

The identical situation was encountered in a nonrenormalizable chiral
theory already at the one-loop level for physical observables
\cite{chiral}. Inside the differential renormalization, one finds no
{\em a~priori} internal criterion for choosing the ratios of auxiliary
masses, to get reliable results. Of course, comparing
eq.~(\ref{beta_diff}) to eq.~(\ref{beta_dim}) in the dimensional
renormalization [or the results for fig.~1(d,e) individually], we can
infer the values that would ensure the invariance:
$\ln(M_2^2/M^2)=\ln(M_3^2/M^2)=1$. But by itself the differential
renormalization remains ambiguous if we apply it to a theory that is not
renormalizable in the ordinary sense, and is not directly compatible
with the background field method.

\section{Conclusion} \indent

Our examples show that the back\-ground-field formalism requires one to
use an invariant renormalization procedure in order to obtain valid
results in a gen\-er\-al\-ized-re\-norm\-al\-iz\-able theory. A
noninvariant regularization or renormalization may break an implicit
correlation between different diagrams, which is essential as one
formally expands the action in the background and quantum fields.

We have demonstrated by direct two-loop calculations that the
regularization via a cut-off in the momentum space is noninvariant and
gives a wrong result for the $\beta$ function of the ${\bf n}$-field
model within the back\-ground-field formalism.

We have also found that the differential renormalization is not
automatically invariant. The result depends on the ratio of the
auxiliary scale parameters beyond the allowed scheme arbitrariness in
the second order of perturbation theory. We can partially fix the
ambiguity by imposing a condition on divergent one-loop tad\-pole-type
diagrams a combination of which should be finite. But this is not
enough, and there seems to be no algorithm of generalizing such
conditions to more complicated graphs.

We would like to stress once more that the calculations in nonlinear
models like the $\sigma $ model or supergravity are hardly possible
without the background-field formalism. Thus, the need in the
regularization that preserves the underline symmetries and
is practically usefull at the same time is of vital importance. The
example considered above clearly demonstrate the problems arising when
using a non-invariant procedure.

\pagebreak[4]

\end{document}